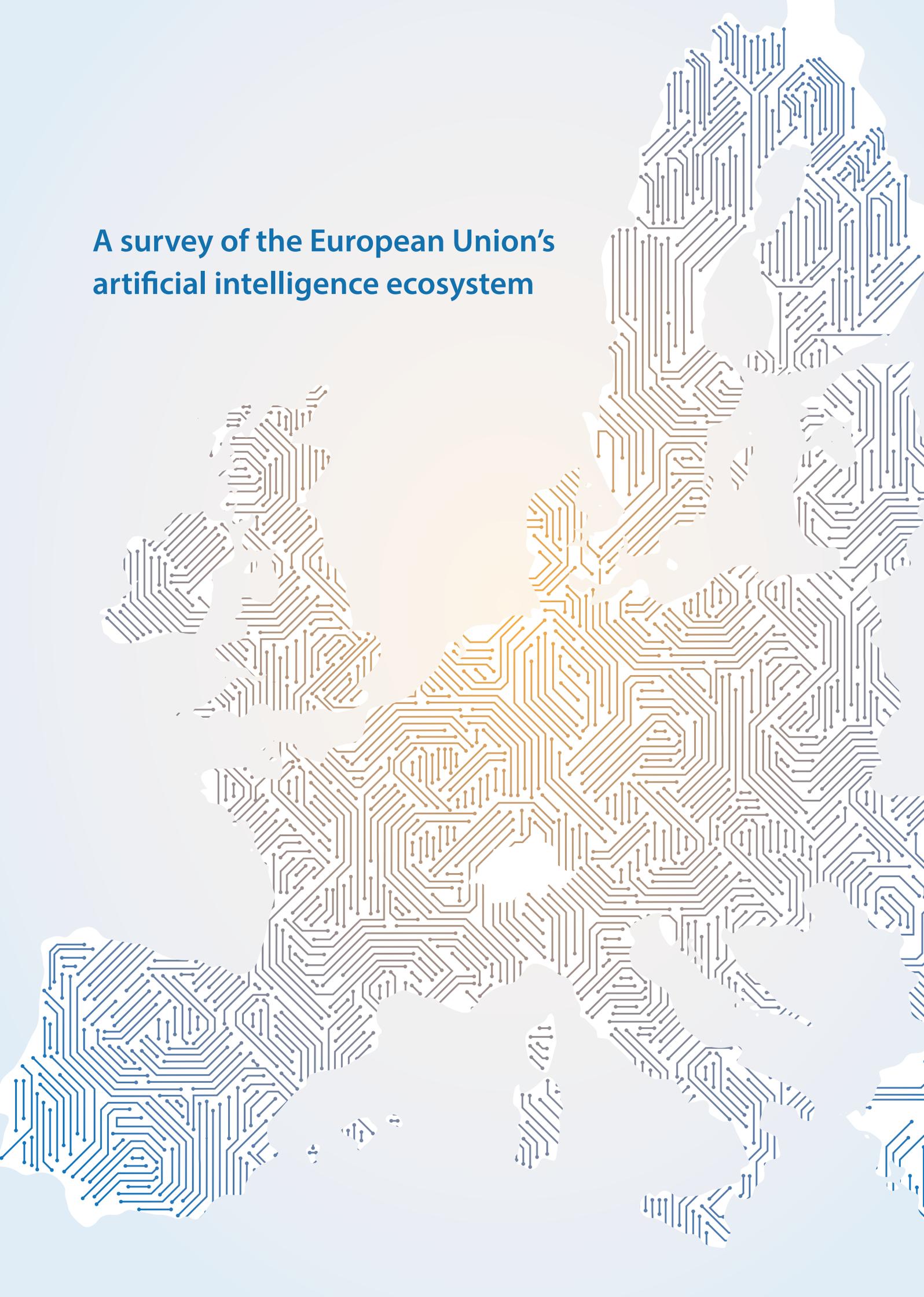

# A survey of the European Union's artificial intelligence ecosystem

Compiled by Charlotte Stix
Research Associate, Leverhulme Centre for the Future of Intelligence, University of Cambridge
Advisor, Element AI

This report is by no means final or comprehensive. It will act as a living document regularly undergoing updates to reflect the rapidly changing AI landscape within the European Union.

Thanks for edits and comments to Seán Ó hÉigeartaigh, Stephen Cave, Matthijs M. Maas, Jess Whittlestone, Emma Bates, Rune Nyrup and Haydn Belfield.

# Contents:

## Summary

## Appendix

# Acronyms:

| | |
|---|---|
| AAAI: | Association for the Advancement of Artificial Intelligence |
| AENEAS: | Association for European NanoElectronics ActivitieS |
| AI: | Artificial Intelligence |
| AI HLEG : | High-Level Expert Group on Artificial Intelligence |
| ARM: | Acorn RISC Machine |
| ARTEMIS: | Advanced Research & Technology for EMbedded Intelligent Systems |
| CERN: | Conseil Européen pour la Recherche Nucléaire (European Organization for Nuclear Research) |
| CIIRC: | Czech Institute of Informatics, Robotics and Cybernetics |
| CLAIRE: | Confederation of Laboratories for Artificial Intelligence Research in Europe |
| DARPA: | American Defense Advanced Research Projects Agency |
| DIHs: | Digital Innovation Hubs |
| ECAI: | European Conference on Artificial Intelligence |
| ECSEL: | Electronic Components and Systems Joint Undertaking |
| EESC: | European Economic and Social Committee |
| EFSI: | European Fund for Strategic Investment |
| EGE: | European Group on Ethics in Science and New Technologies |
| EIB: | European Investment Bank |
| EIF: | European Investment Fund |
| ELLIS: | European Lab for Learning & Intelligent Systems |
| EPoSS: | European Platform on Smart Systems Integration |
| EU: | European Union |
| EurAI: | European Association for Artificial Intelligence |
| GDPR: | General Data Protection Regulation |
| HBP: | Human Brain project |
| EuroHPC JU: | High-Performance Computing Joint Undertaking |
| HQ: | Headquarters |
| H2020: | Horizon 2020 |
| ICT: | Information and Communications Technology |
| IEEE: | The Institute of Electrical and Electronics Engineers |
| IJCAI: | International Joint Conference on Artificial Intelligence |
| JEDI: | Joint European Disruptive Initiative |
| MAR: | SPARC's Multi-Annual Roadmap |
| MILA: | Montreal Institute for Learning Algorithms |
| MOOC: | Massive Open Online Course |
| MFF: | Multiannual Financial Framework |
| NATO: | North Atlantic Treaty Organization |
| PPP: | Public-Private Partnership |
| PRAIRIE: | Paris Artificial Intelligence Research Institute |
| R&D: | Research and Development |
| R&D&I: | Research, Development and Innovation |
| SMEs: | Small and Medium sized Enterprises |
| SPARC: | Public-Private Partnership for robotics in Europe |
| SRA: | SPARC's Strategic Research Agenda |
| STEM: | Science, Technology, Engineering and Math |
| S&T: | Scientific and Technological |
| VC: | Venture Capital |
| WG: | Working Groups |

# Executive Summary

Narratives in international media, and increasingly within governments, place great importance on nations achieving leadership in artificial intelligence (AI). The EU[1,2] is rarely considered the leading player in these discussions.[3] This report investigates this assumption and outlines existing building blocks that could form the basis for EU leadership in AI. The research is based primarily around EU legislation, policy and strategy documents, publicly available databases of ongoing projects, and funding decisions. In aggregating this information, the report aims to provide an introductory overview of the EU's AI ecosystem.

The report is structured into the following sections: (1) strategy and vision[4,5] and (2.a) funding and financial support, (2.b) talent creation [6,7] (2.c) collaboration.[8] The primary conclusions around each of these are as follows:

(1) The EU, via the European Commission in their Communication on AI, sets out a "European Initiative on AI" as well as a Coordinated Plan on AI from the EU Member States.[9] This initiative is supported with a declaration signed by all 28 Member States, the Declaration of Cooperation on AI. The Coordinated Plan on AI outlines how EU Member States could coordinate their strategies, financial commitments and other resources to increase European competitiveness as a whole. Perhaps most notably, the EU's vision for AI is centered around 'ethical AI' in a way that could distinguish it from its American or Chinese counterparts.

(2.a) Although it is too early to judge their impact, the EU is undertaking active steps to tackle funding bottlenecks.[10] The European Commission will invest €1.5bn[11] between 2018-2020 towards research and innovation in AI, a 70% increase from current investments, and expects to invest €2.5bn[12] during the Digital Europe Programme (2021-2027) alone. These numbers exclude funding from other sources such as the European Research Council and the European Fund for Strategic Investment. The EU is struggling to attract venture capitalists. They invested only €6.5bn in the EU in 2016, in comparison to €39.4bn in the US.[13] Newly-established initiatives such as VentureEU aim to redress this imbalance.

(2.b) The EU has a number of initiatives dedicated to talent creation for the ICT and digital sector. However, there is mounting concern over brain drain and a loss of future-oriented entreprises, with researchers leaving the EU and a number of promising companies being acquired by international companies, such as Acorn RISC Machine (acquired by SoftBank),[14] KUKA (Midea),[15] and Magic Pony Technology (Twitter)[16]. Talent creation through education, re- and upskilling may put the EU on a solid foundation for future AI competitiveness, but brain drain remains a substantial concern.

(2.c) The EU has several ongoing and upcoming initiatives between Member States and groups within those that are of collaborative nature. Large-scale infrastructural collaborations such as the Digital Innovation Hubs and the AI4EU pilots exist alongside prominent collaborations on resources, evidenced by the Joint Undertakings for electronic components as well as high-performance computing. Looking forward, the EU could build on a track record of major collaborative projects such as the Human Brain Project and CERN, which could provide a model for collaborative AI initiatives of similar scale.

This report does not judge whether these components will suffice for the EU to achieve or maintain AI leadership. Future work will be needed to assess the effectiveness of these initiatives and compare them to those of other nations. However, by providing an overview of the EU AI ecosystem this report hopes to open a broader discussion about what EU leadership in AI could look like and highlight many of the components that could provide its basis.

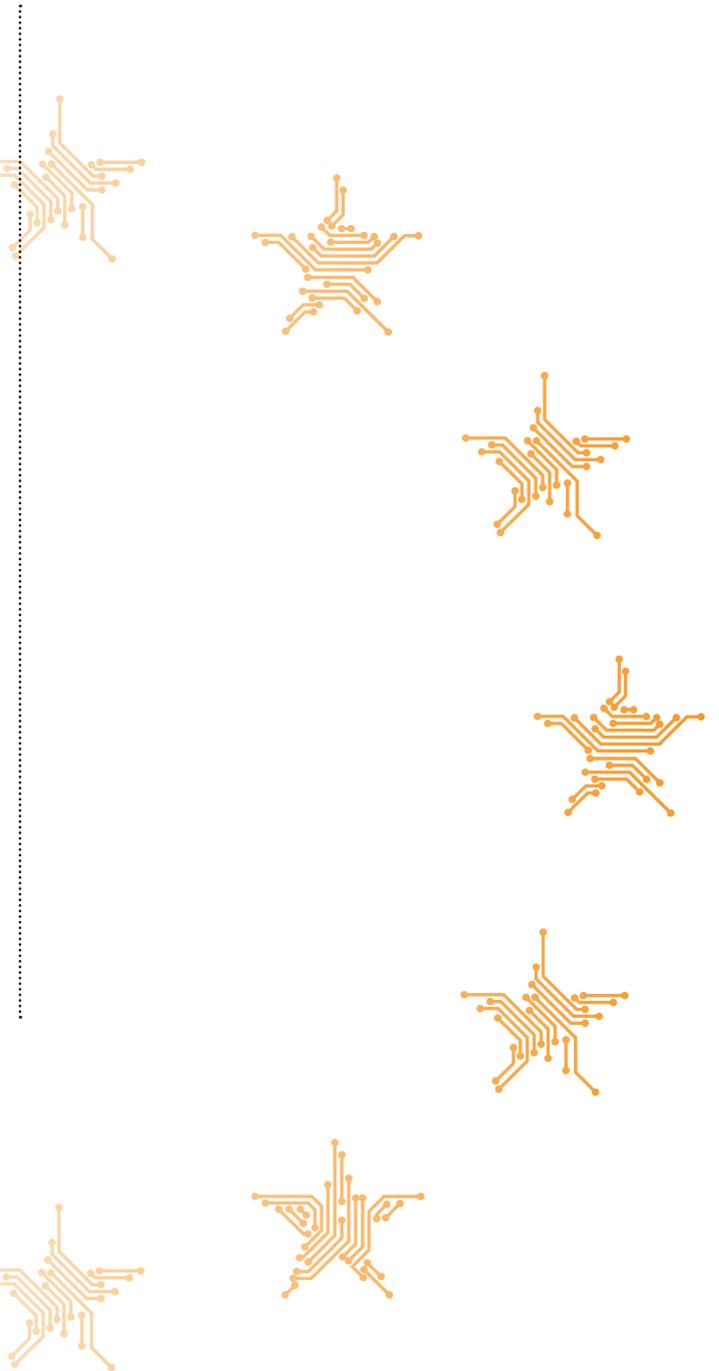

# Introduction

Recent progress in AI has led a variety of countries to set up governmental advisory boards, propose AI strategies and establish new institutions that focus on the societal impact of this technology. Some countries and strategies feature more prominently in public fora than others. This has the potential to contribute to a misleading impression of the shape and state of the global AI ecosystem. In contrast to other global powers, the EU appears at times reluctant to trumpet achievements, initiatives or ambitions, be that in AI or other fields. As a result, awareness about the work undertaken by the EU is often lacking, which, in turn, contributes to the narrative that it is on the waiting bench when it comes to AI.

This report aims to rectify this impression by providing a deeper insight into the EU's AI ecosystem through an investigative tour of past, current and proposed AI-relevant activities. Several areas are examined in a closer manner: (1) strategy and vision,[17,18] (2.a) funding and financial support, (2.b) talent creation,[19,20] and (2.c) collaboration.[21]

Part 1 outlines the EU's strategy and vision with regard to AI. It is divided into two sections. Section (a) looks directly at the EU's AI strategy. Section (b) explores the accompanying vision.

Section (a) outlines the EU's AI strategy through the introduction of three key publications: the Digital Day Declaration **'Declaration of Cooperation on AI'**, the European Commission's **'Communication on Artificial Intelligence'** and the European Commission's **'Coordinated Plan on AI'**. It then presents the two groups that help steer the European Commission in the implementation of several components of these strategy documents: the **High-Level Expert Group on Artificial Intelligence** and the **European Artificial Intelligence Alliance**. The first group is composed of 52 subject experts who act as an advisory group to the European Commission on ethics, investment and policy-making and work on outlining a long-term AI strategy. The second group is a multi-stakeholder forum open to all EU citizens to share their opinion, concerns and feedback on developing AI policy and strategy. Crystallising out of section (a) is the image of the EU as an actor that self-identifies as a leader in 'ethical AI'.[22]

Section (b) explores this on the basis of recent reports, papers, regulations and EU-funded projects. It introduces three relevant documents, the European Group on Ethics in Science and New Technologies' statement on **'Artificial Intelligence, Robotics and 'Autonomous Systems"**, the High-Level Expert Group's **'Draft Ethics Guidelines on AI'**, and the European Economic and Social Committee's **'Opinion on Artificial Intelligence'**.[23] All three publications demonstrate the importance that actors within the EU place on human-centric and 'ethical AI'. This is followed by a brief examination of regulations that could contribute to 'ethical AI', such as the **General Data Protection Regulation** and **ePrivacy Regulation**. Finally, two EU funded projects are discussed, concerning related issues such as data usage, ownership and algorithmic decision making from a non-regulatory perspective: **DECODE** and **Algo Awareness**. Part 1 concludes with a summary of the EU's AI strategy and vision.

Part 2 then investigates the granular elements that will support the EU in the implementation of its strategy and vision. It is divided into three sections: (a) funding and financial support, (b) talent creation and (c) collaboration.

Section (a) explores funding and financial support. It looks at the newly proposed **VentureEU** fund and the **European Fund for Strategic Investment**. It then broadly covers other funding sources for research, development and innovation such as the **Horizon 2020** framework programme for research and innovation (2018-2020), **Horizon Europe** (2021-2027) succeeding Horizon 2020, the **European Innovation Council** and the upcoming **Digital Europe Programme**.

Section (b) explores talent creation. The EU tackles talent creation through education, upskilling and reskilling, as evidenced by e.g the **New Skills Agenda** and the **Digital Skills and Jobs Coalition**. Various national strategies, such as Germany's *Eckpunkte der Bundesregierung für eine Strategie Künstliche Intelligenz*,[24] Finland's *Age of Artificial Intelligence*[25] and France's *Donner un sens à l'intelligence artificielle*[26] propose more direct next steps to halt brain drain.

Section (c) investigates areas of collaboration and cooperation. It is divided into two subsections: (i) exploring initiatives at macro-level (i.e. collaborations between Member States) and (ii) exploring initiatives at a more granular level. Subsection (i) highlights existing collaborative infrastructure: the **AI4EU project** (formerly AI-on-demand platform) and the **Digital Innovation Hubs** initiative. Both aim to increase access to AI, including to data, tools, and advice, for small and medium sized enterprises (SMEs). This is followed by the introduction of a number of other EU-wide collaborations that are likely to play an important role for the EU's AI leadership. In particular, it outlines **SPARC**, a Public-Private Partnership between the European Commission and the robotics community, the **Electronic Components and Systems Joint Undertaking** and the **European High-Performance Computing Joint Undertaking**.

The latter two both focus on AI-relevant hardware. It ends with the introduction of three eminent proposals for a European AI laboratory: **CERN for AI**, the **European Lab for Learning & Intelligent Systems** and the **Confederation of Laboratories for Artificial Intelligence Research in Europe**.

Subsection (ii) explores more granular collaborative actions and initiatives. The **Joint European Disruptive Initiative** is a Franco-Germanic initiative that aims to redress the EU's lack of funding for 'moonshot' projects. The **Paris Artificial Intelligence Research Institute** is a proposed initiative to develop a distributed AI research "institute of institutes" across Paris with links to global AI labs. The report then focuses on academic collaborations by introducing **EurAI**, the European Association for AI. In light of the identified initiatives, community movements and ongoing networks, Part 2 concludes with a summary of current opportunities as well as drawbacks.

The report ends with two main conclusions. The first is that the EU has a suitable and encouraging framework to become a leader in 'ethical AI'. This includes regulations, citizen engagement and the development of ethics guidelines.[27] While leadership in 'ethical AI' is important, it must be incorporated within the broader goals of an EU AI strategy and the proposed implementation process of the European Commission's Coordinated Plan on AI.[28] This will likely require continued support for cutting-edge research capability in AI and for successful commercial applications of AI as well as the meaningful alignment of national AI strategies. The second conclusion is that the EU should pay diligent attention to the funding for start-ups and talent brain drain while the impact of newly established and proposed initiatives remains unclear.

The annex provides a brief overview of existing large-scale pan-European collaborations. These are not directly related to AI, but provide a model for what the EU could aim for when it comes to large-scale AI research: **CERN** and the **Human Brain project**.

# Part 1: The overarching plan

Part 1 outlines the EU's overarching strategy and vision. It is divided into two sections: (a) looks at recent developments with regard to the EU's AI strategy, and presents the EU's intention to be a leader in 'ethical AI'. Section (b) supports the picture of the EU as a viable leader in 'ethical AI' on the basis of recent reports, papers, regulations and relevant EU projects. Subsequently, part 2 places this general strategy within the broader ecosystem and explores areas of collaboration, funding and talent creation.

## Section (a): The European Union's AI strategy

In 2018, the European Commission presented the Digital Day Declaration 'Declaration of Cooperation on AI',[29] now signed by all 28 Member States (plus Norway). In the Declaration,[30] Member States agree to stay in close dialogue with the European Commission and to work together towards "a comprehensive and integrated European approach on AI and, where needed, review and modernise national policies to ensure that the opportunities arising from AI are seized and the emerging challenges addressed."[31]

The main goal of the Declaration is to build a framework for cooperation between Member States on areas ranging from AI's impact on the labour market, sustainability and trustworthiness to ethics and funding. Furthermore, the Declaration draws attention to pressing challenges in areas such as education and reskilling. Other points deal with, for example, the importance of cooperation when it comes to the expansion of and support for AI research centres, a salient point related to the discussions in Part 2. It also touches upon the mitigation of arising legal, ethical and socio-economic risks.

In fact, the Declaration demonstrates a clear concern for the ethical issues arising out of current and future AI-system's development and deployment. Following this concern, it commits its signatories to ensuring that 'humans remain at the centre of AI development', and to prevent the "harmful creation and use of AI applications".[32,33]

Despite its non-binding nature, the Declaration should be seen as a serious demonstration of intent on behalf of the Member States to collaborate on AI and to strengthen EU leadership. An intent taken up from the European Commission's 'Communication on AI' and further sustained by the subsequent 'Coordinated Plan on AI'.

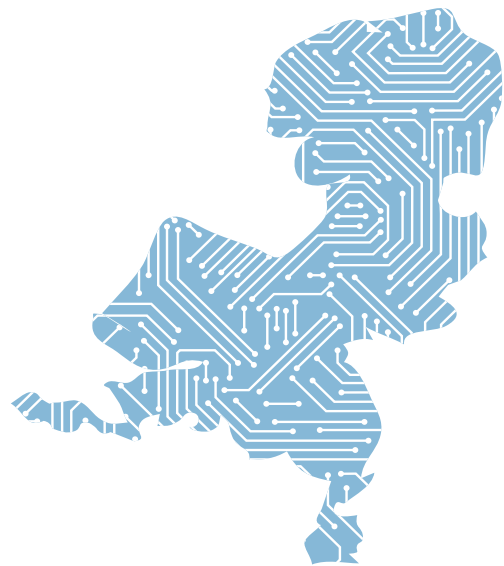

One of the two core strategy documents of the European Commission is the 'Communication on AI',[35] published in response to the European Council's call to put forward a European approach to AI. The Communication effectively lays the groundwork for the 'Coordinated Plan on AI' and outlines the steps needed to achieve a more committed alignment of resources and goals between Member States. The whole, it appears, is expected to be greater than the sum of its parts. To that end, the Communication states that the EU "should have a coordinated approach to make the most of the opportunities offered by AI and to address the new challenges that it brings".[36]

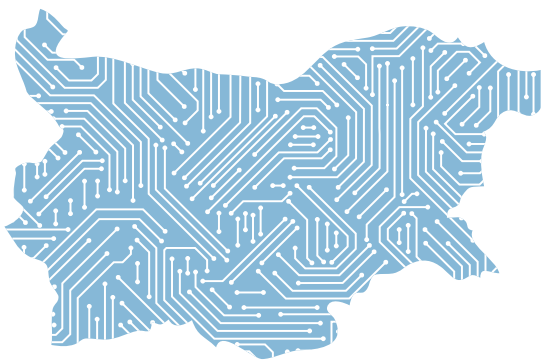

It further suggests three core steps to strengthen the current EU AI ecosystem: 1. boost Europe's technological and industrial capacity as well as AI uptake; 2. prepare Europe for the socio-economic changes associated with AI; 3. ensure that Europe has an appropriate ethical and legal framework to deal with AI development and deployment. The Communication proposes several financial and infrastructural mechanisms to increase investment for stagnating areas such as funding into startups and business access to AI tools.[37] Other projects of note are the development of regulatory sandboxes,[38] a commitment to support centres for data sharing and to update access to and preservation of scientific information.

The 'Coordinated Plan on AI'[40] is the European Commission's second strategy document. It's a non-binding proposal building on the previously published 'Communication on AI' and the Declaration of 'Cooperation on AI'. The Plan includes a projected aim of €20bn in funding by 2020, with gradual target of €20bn on a yearly basis thereafter, and lays the foundation for coordination on AI between Member States and other stakeholder groups, with an invitation to cooperate with international stakeholders sharing the same values. This could be considered as the first case of a set of countries attempting a light-touch form of coordination on AI governance.

The Plan's overarching rationale is that coordination between the Member States can increase the EU's global competitiveness by maximising investment on EU and Member State level, encouraging synergies between ongoing efforts (incl. ethics), exchanging best practices and "collectively define a way forward"[41], i.e. creating a common goal and vision. The Plan outlines a European approach to AI that is built upon ethical and societal values derived from the Charter of Fundamental Rights. It places an emphasis on what it perceives to be interconnected concepts of a 'trusted AI' and 'human-centric AI'. Key principles identified for the broader goal of achieving AI made in Europe include 'ethics by design' and 'security by design'. It states that "overall, the ambition is for Europe to become the world-leading region for developing and deploying cutting-edge, ethical and secure AI, promoting a human-centric approach in the global context."[42] To support this, safety and liability frameworks will be assessed in light of adequate safety and redress mechanisms and, more generally, regulatory frameworks will be assessed for fitness of purpose in regard to AI-enabled technologies.

The implementation of the Plan will be supported by the Member States' Group on Digitising European Industry and Artificial Intelligence, steering conversations between

Member States and the European Commission. There will be bi-annual meetings as well as coordination actions across national ministries, industry, academia, civil society and other stakeholders. In addition, academia, industry and the European Commission, with the support of Member States, will work on a common Research and Innovation Agenda for AI (2020).

In terms of clear immediate financial goals, Horizon 2020 funding for AI will increase by 70% between 2018-2020, totalling at €1.5bn. Following, the European Commission proposes an investment of €1bn per year under the next Multiannual Financial Framework (MFF), drawing on funds from Horizon Europe and the Digital Europe Programme. The Coordinated Plan on AI also reiterates the 'Communication on AI''s ambition to increase investment to reach a total of €20bn until 2020, including public and private funding.

The following areas are identified as being in particular need of coordination: investment, "excellence in and diffusion of AI",[43] data availability (pool resources such as data), societal challenges, ethics and regulatory frameworks.

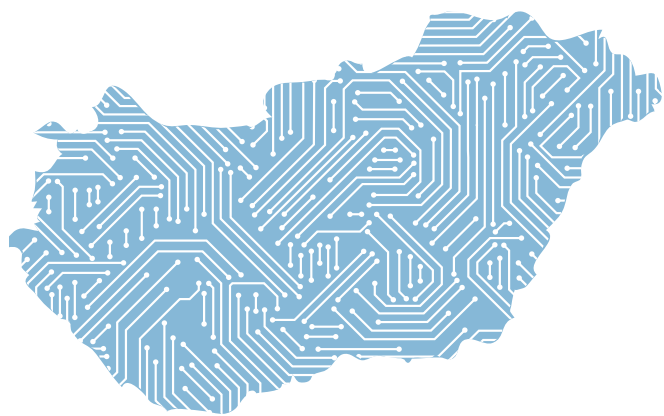

### Coordinated Plan on the Development and Use of Artificial Intelligence Made in Europe

Published: 7th December 2018

Author: European Commission

Objective: "To make these efforts a success [...] Member States and the Union should attempt to align bilateral outreach efforts related to AI between individual Member States and third countries and pool their efforts pushing for a responsible development of AI at the global level. The Union needs to speak with one voice to third countries and the world at-large on this topic. In synergy with activities of the Member States, the EU should also seek alliances with stakeholders —tech companies, academia and other parties— to engage in a multi-stakeholder alliance at the global level for responsible AI."[44]

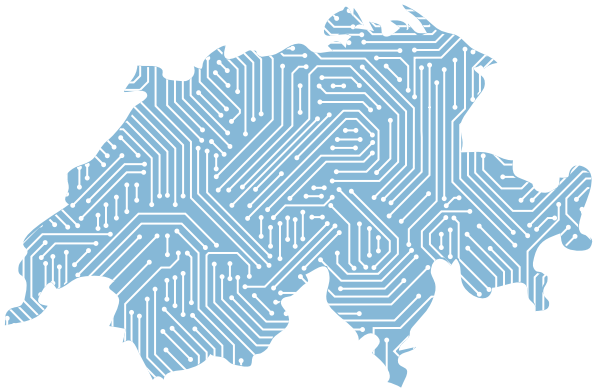

In order to support the development of the AI strategies outlined above, the European Commission established two groups, as described in the Communication on AI: the High-Level Expert Group on AI (AI HLEG)[45] and the European AI Alliance.[46]

The High-Level Expert Group on AI contributes to the shaping of the EU's AI strategy, policy and priorities. It advises the European Commission on both near- and longer-term challenges as well as opportunities arising from AI. The AI HLEG further acts as the steering group to the European AI Alliance. Its experts were chosen through an open call by the European Commission, with the final Group comprising 23 members from industry, 19 from academia, and 10 from civil society. It organises and manages internal and external multi-stakeholder dialogues, resulting in reports and policy suggestions for the European Commission.

The AI HLEG is divided into two working groups (WGs): one on ethics and one on investment and policy. The WG on ethics' remit is to draft AI ethics guidelines.[47] This includes an examination of the impact of AI on the Charter of Fundamental Rights as well as addressing considerations such as non-discrimination, dignity, privacy and personal data protection, safety and transparency. They recently published their first draft for public consultation through the AI Alliance. The final Guidelines shall be presented to the European Commission by Q1 2019. The WG on investment and policy is tasked with the development of policy and investment recommendations to aid the European Commission in achieving the goals described in the Communication on AI. This includes measures to strengthen the EU's competitiveness, guidance for the Strategic Research Agenda on AI and the establishment of a pan-European network of AI labs.

> **High-Level Expert Group on AI**
>
> Established: 14th June 2018
>
> Members: 52 members from civil society, academia and industry[48]
>
> Objective: "1. Advise the Commission on next steps addressing AI-related mid to long-term challenges and opportunities [..]; 2. Support the Commission on further engagement and outreach mechanisms [..]; 3. Propose to the Commission AI ethics guidelines [..]."[49]

The second group, the European AI Alliance, is built around a diverse multi-stakeholder online platform. There, members can contribute to ongoing discussions on AI, feeding into the European Commission's policy-making. They can also directly engage with the AI HLEG, which publishes its draft suggestions on the platform for feedback. The European AI Alliance is open to all members of society. Currently, it is composed of civil society, members of trade unions, companies, not-for-profit institutions and consumer organisations. The Alliance is ambitious in intent; however, the sheer number of members might eventually complicate the platform's management, dilute reasonable voices and ultimately decrease its impact. At the same time the European AI Alliance represents a strong commitment to and leading attempt[50] at an actual cross-societal and pan-European multi-stakeholder dialogue.

Together, the elements mentioned above form the broad outline of the EU's current strategy for AI. The tonal quality thereof and repeated intention to create 'human-centric', 'trustworthy AI' and 'ethical AI' suggests the EU envisions itself as a leader in these areas.

## Section (b): The European Union's Vision

Section (b) expands on the EU's vision of itself as a leader in 'ethical AI'. When it comes to the creation of a framework for 'ethical AI' the EU may benefit from its "unity in diversity",[51] allowing it to draw on various cultural and historic backgrounds. The EU's path towards AI is one that seems to align with its historic values, including the EU's fundamental values[52] as defined in the Treaty of Lisbon and its adherence to the European Convention on Human Rights.[53] Echoing this, the European Commission's Communication on AI states that: "the EU must therefore ensure that AI is developed and applied in an appropriate framework which promotes innovation and respects Union's values and fundamental rights as well as ethical principles such as accountability and transparency. The EU is also well placed to lead this debate on the global stage. This is how the EU can make a difference - and be the champion of an approach to AI that benefits people and society as a whole."[54]

Beyond the commitment to a diverse and inclusive dialogue as demonstrated by the AI Alliance, there are several recent publications,[55] regulations and projects that strengthen this self-affirming vision of the EU. The most directly relevant publications are the European Group on Ethics in Science and New Technologies' (EGE) statement on 'Artificial Intelligence, Robotics and 'Autonomous Systems",[56] the AI HLEG's 'Draft Ethics Guidelines for 'Trustworthy AI", and the European Economic and Social Committee's (EESC) 'Opinion on Artificial Intelligence, Artificial intelligence – The consequences of artificial intelligence on the (digital) single market, production, consumption, employment and society '.[57]

The EGE's statement calls for the establishment of an overarching framework for AI. To that end, it proposes the development of several ethical principles based on fundamental European values. The principles should act to tackle legal, ethical and governance issues arising from AI development and deployment. Ultimately, their goal should be to ensure that AI is created with "humans in mind". In a forward-thinking manner, the statement also considers that future developments in AI and robotics may need to be accompanied with the introduction of several new rights, such as a "right to meaningful human contact and the right to not be profiled, measured, analysed, coached or nudged".[58]

To ensure regulatory and ethical foresight, the EGE's statement suggests that a better understanding of future AI-based technologies is needed. It claims that forecasting and measurement in this area could contribute to more accurate policy making by governments and benefit the EU. It is not alone in pointing towards the importance of measurement: the Communication on AI mentions the importance of technology monitoring, and national strategy documents such as Germany's *Eckpunkte der Bundesregierung für eine Strategie Künstliche Intelligenz* propose AI monitoring as a method to produce AI-relevant seals of quality and to set standards. The EU, with a tradition of high-quality standard setting, for example with the European Emission Standards,[59] and certification, for example the Certification Europe (CE)[60] certifications, could occupy a leading role in AI measurement, standardisation and forecasting. Its attempt to position itself as leader in 'ethical AI' could provide additional credibility for the setting of standards in the areas of safety and human-centric AI development.

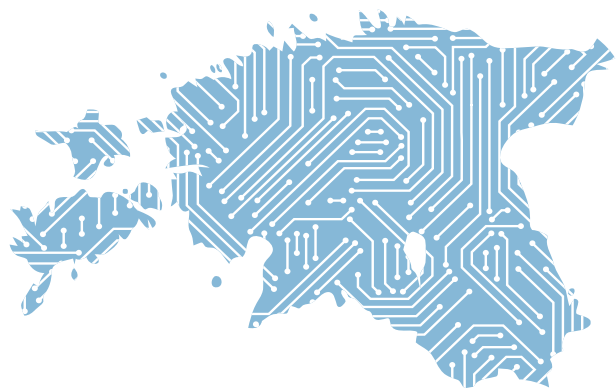

The statement also highlights the lack of coordinated, overarching regulations for AI and AI-based technologies in the EU as a possible blindspot. Potential downsides include reduced competitiveness and 'ethics shopping' (i.e. a situation where companies relocate to or conduct business in countries with lower ethical standards).

> **Artificial Intelligence, Robotics and 'Autonomous Systems'**
>
> Published: 9th March 2018
>
> Author: European Group on Ethics in Science and New Technologies, European Commission (EGE)
>
> Rapporteur: Jeroen van den Hoven
>
> Ideas: Among others, the statement suggests ideas for a set of ethical guidelines and democratic prerequisites. They centre around data protection and privacy, sustainability, rule of law and accountability, security, safety, bodily and mental integrity and democracy.

The AI HLEG's 'Draft Ethics Guidelines for 'Trustworthy AI''[61] are the clearest indicator of the EU's ambition to become a leader in 'ethical AI'. Although they solely demonstrate the expert group's opinion and are not an official European Commission document, the guidelines nonetheless constitute an insight into the direction that Europe is heading in, namely towards 'human-centric and trustworthy AI'. The draft document contains a variety of chapters ranging from AI's compliance with fundamental rights, technical and non-technical methods to implement the outlined 'trustworthy AI' to an assessment list and case studies for 'trustworthy AI'. 'Trustworthy AI' as used by the AI HLEG is described as made up of two components: "(1) it should respect fundamental rights, applicable regulation and core principles and values, ensuring an "ethical purpose" and (2) it should be technically robust and reliable since, even with good intentions, a lack of technological mastery can cause unintentional harm."[62] The overall ambition of the guidelines appears to be to incite a discussion of ethical frameworks for AI "at a global level", one where Europe's approach of using ethics as "inspiration to develop a unique brand of AI"[63] is seen as it taking a leadership role and crucial to enable its competitiveness. Furthermore, it is expected to serve as as starting point for a European discussion on "Trustworthy AI made in Europe".[64] Given that this is only a draft document and will undergo significant changes after the stakeholder consultation is over, it is unclear how the final guidelines will look like.

> **Draft Ethics Guidelines for Trustworthy AI**[65]
>
> Published: draft published 18th December 2018, undergoing stakeholder consultation
>
> Authors: the AI HLEG's WG on ethics with input from the whole group
>
> Objective: "Given that, on the whole, AI's benefits outweigh its risks, we must ensure to follow the road that maximises the benefits of AI while minimising its risks. To ensure that we stay on the right track, a human-centric approach to AI is needed, forcing us to keep in mind that the development and use of AI should not be seen as a means in itself, but as having the goal to increase human well-being. Trustworthy AI will be our north star, since human beings will only be able to confidently and fully reap the benefits of AI if they can trust the technology."[66]

Although the EESC's[67] Opinion on AI[68] is the earliest document published from the selection observed, it already hints at narratives recurrent in the above mentioned documents. It recommends the development of a standardisation system that could verify, validate and monitor AI and AI-based systems and advises that the EU ought to establish a clear policy framework for AI, to ensure global leadership. Both recommendations are salient and strongly dependent on the EU's ability to work as a cohesive whole when it comes to aligning their policies and strategy.

Overall, the Opinion on AI touches on a variety of elements crucial to AI development and deployment, with a heavy focus on ethical, societal and general safety considerations. One recommendation, aligned with many other EU documents and tackled in the Draft Guidelines, is the development of a code for ethics (for the entire creation process from design to development to deployment).[69] It further backs the call for a ban on lethal autonomous weapons systems and briefly delves into a consideration of the dangers that artificial superintelligence could pose.[70]

In a way, the Opinion can be seen as one of the inflection points for the EU's vision of itself as leader in 'ethical AI'. Throughout the document the rapporteur advocates for a "human-in-command" approach, urges Europe to support and promote the "development of AI applications that benefit society"[71] and even provides tangible next steps to do so.

> **Opinion on AI: Artificial intelligence – The consequences of artificial intelligence on the (digital) single market, production, consumption, employment and society**
>
> Published: 31st August 2017
>
> Rapporteur: Catelijne Muller
>
> Objective: The EESC published recommendations concerning the following 11 areas in which AI may create societal challenges: "ethics; safety; privacy; transparency and accountability; work; education and skills; (in)equality and inclusiveness; law and regulations; governance and democracy; warfare; superintelligence."[72]

As demonstrated above, the EU puts a clear emphasis on human-centric and ethical AI, which supports the claim that it sees itself both as a leader in these areas and, related, indicates that it takes this as distinguishing element when compared to other powers aiming for AI leadership.

This emphasis is further supported by recent EU-wide regulation and ongoing projects which the remainder of this section examines. In particular, it will focus on the usage of algorithms and data as well as on citizen empowerment as an avenue towards 'ethical AI'.

When it comes to data protection, the General Data Protection Regulation (GDPR) is a prime example. Discussed for six years,[73] it took around two years to be finally approved by the European Parliament, replacing the 1995 Data Protection Directive.[74] While a directive leaves it open to the Member States how to transpose it into national law (as long as the end goal is reached), a regulation is immediately applicable and enforceable. This means that it creates a single approach applicable to[75] and enforceable by all actors and Member States within the EU.[76] In this regard, the GDPR contributes to a cohesive regulatory framework for data protection and privacy across the EU.

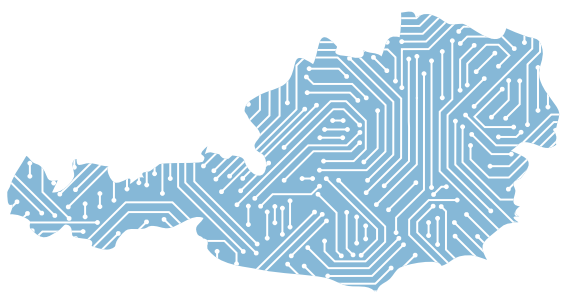

In terms of content, the GDPR grants EU citizens greater agency over the way their personal data is being used and reused, and aims to protect it against potential misuse. As part of this, EU citizens have gained a host of new rights: e.g. a right to be forgotten, a right to access and data portability and to privacy by design. The right to be forgotten effectively endows the citizen[77] with the right to withdraw consent. This means they can demand that previously provided personal information be erased and no longer shared, including with third parties. The right to access grants the citizen with the right to demand a copy of their personal data held by the relevant organisation.[78] Data portability endows the citizen with the right to move previously provided personal data from one organisation to another. Finally, privacy by design entails that data protection must be part of the initial design of a system.

As a complementary piece of legislation, the EU will soon introduce a new ePrivacy regulation.[79,80] This would expand and clarify the GDPR's reach as well as broaden its scope to cover areas under online communication.

Despite potential shortcomings, the GDPR is a clear commitment to strengthen data privacy, protect and empower users and by extension lay the foundation for a regulatory approach towards AI that is human-centric.[81]

> **General Data Protection Regulation**
>
> Came into force: 25th May 2018
>
> Applicable to: all relevant actors within the EU and those who process, monitor or hold data on EU citizens outside of the EU.
>
> Objective: strengthen EU citizens' control over their data and protect them from data and privacy breaches.

In addition to regulatory action, there are also initiatives exploring data usage for the public good such as the DEcentralised Citizen-owned Data Ecosystems (DECODE).[82] DECODE's aim is to empower citizens to proactively chose the purpose for which their data is used and the method by which it can benefit the public good. At the core of DECODE is a digital wallet, based on distributed ledger technology. The wallet allows each participant to manage if, how and with whom they wish to share their personal data. The idea is not completely unlike the e-Estonia model, where citizens own their information and can access it through a secure online platform.[83] DECODE currently runs four pilot programs. Each program allows citizens to decide which platforms can access their data. Simultaneously, citizens allow government to use their shared data for public use, for example through the establishment of data commons.[84] The pilot programs cover areas like healthcare, where citizen's data is shared with the government to inform pertinent policy, and social engagement, where data is used to establish neighbourhood social networks.

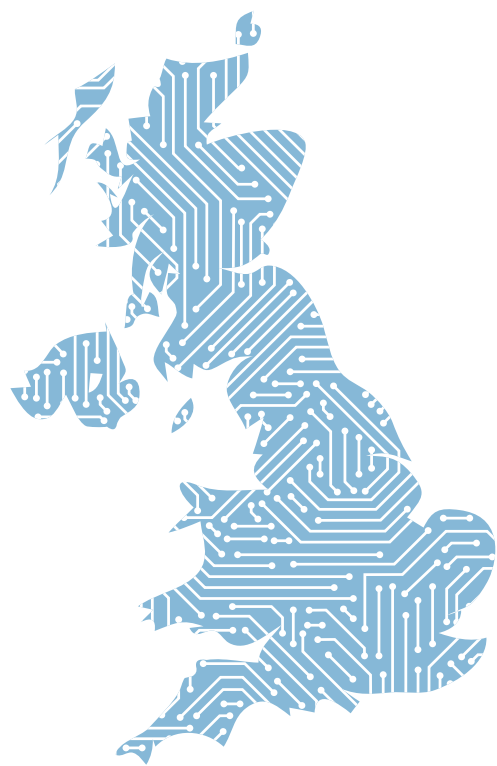

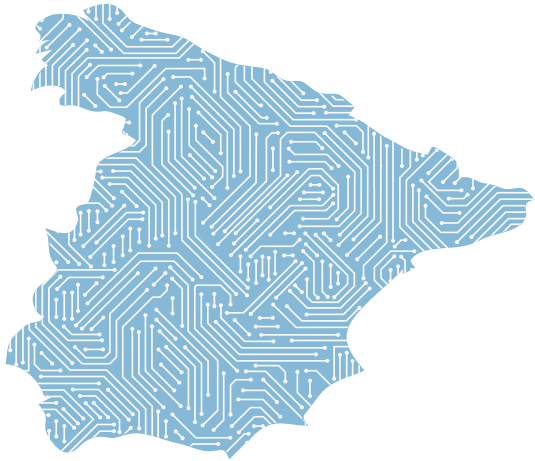

and shortcomings associated with algorithmic decision-making. This is expected to help raise public awareness, and contribute to the creation of suitable policy measures. Together with ongoing projects like DECODE and existing regulation such as the GDPR, it can create a landscape where citizens comprehend and steer ongoing debates surrounding data and algorithms.

> **Algo Awareness**
>
> Established: pilot phase
>
> Objective: "1. contributing to a wider, shared understanding of the role of algorithms, particularly in the context of online platforms, with the intention of raising public awareness and debate of emerging issues; 2. identifying the types of problems, emerging issues and opportunities raised by the use of algorithms, and establish a scientific evidence-base for these issues and opportunities; and 3. designing and prototyping a policy toolbox including solutions for a selection of problems, including policy options, technical solutions and private sector and civil society-driven actions."[86]

## Conclusion

The EU's budding strategy and vision for AI, through the strategic documents and the two main steering groups discussed, unearths a recurrent narrative of 'human-centric' and 'ethical' AI. This naturally leads to a conclusion that the EU envisions itself as a leader in the broad field of 'ethical AI'. This claim is further evidenced by recently introduced regulatory measures (GDPR and ePrivacy Directive), publications (see above) and complementary projects (DECODE and Algo Awareness). Throughout Part 1 it has become evident that the EU's actions align with the vision it has set for itself as a leader in 'ethical AI'. This ambition may contribute to a unique competitive advantage for the EU in the AI ecosystem,[87] if harnessed quickly and appropriately.

> **DECODE**
>
> Funded: European Commission with close to €5m
>
> Projects: Four Pilot Projects
>
> Members of Pilot Projects: Spain (Barcelona), Netherlands (Amsterdam)
>
> Consortium: 14 partners from across Europe: BCMI Labs AB, City of Amsterdam, CNRS (Centre d'économie de la Sorbonne), Dyne, Eurecat, Institut Municipal d'Informatica de Barcelona (IMI), Nesta, Open University of Catalonia, Politecnico di Torino/Nexa, Stichting Katholieke Universiteit Nijmegen Privacy & Identity Lab, Thingful, ThoughtWorks Ltd, University College London, Waag Society.

Another relevant project, currently in its conception phase, is Algo Awareness[85] which is expected to explore best practices to mitigate risks arising out of algorithmic decision-making. To that end, it will analyse potential risks and conduct a study with citizens, focussing on understanding the impact algorithmic-decision making can have on information flow. It will also investigate how algorithmic decision-making can impact citizens' lives. The analysis and study will result in a reference base outlining benefits

# Part 2: The ingredients

## Section (a): Funding and financial support

The EU's competitiveness is particularly hampered by a lack of VC investment and funding for startups.[88] According to a recent European Commission press release,[89] the EU had an investment of about €6.5bn by venture capitalists in 2016, whereas in the US that number reached €39.4bn.[90] While more general funding programmes and national funding vehicles[91] exist, they are often difficult to locate and limited in size and scope. [92]

The EU cannot retroactively undo past dynamics and oversights, but it is beginning to address these challenges going forward. In addition, the landscape appears to be slowly changing, with VC investment in Europe at a steady rise.[93]

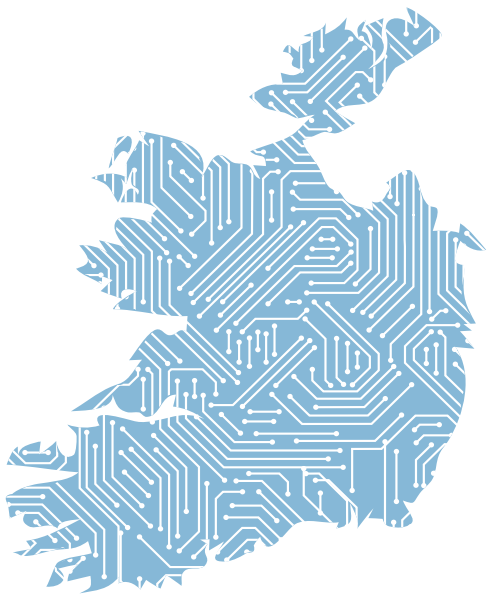

The European Commission in partnership with the European Investment Fund set up a pan-European VC Funds-of-Funds programme, VentureEU, to tackle the current lack of private investment. VentureEU[94] benefits from €410m which it will distribute to startups and companies that look to scale up within the area of AI. It is furthermore expected to raise an additional €2.1bn from public and private investment.

> **VentureEU**
>
> Established: 2018
>
> Lead by: European Commission and European Investment Fund
>
> Budget: €410m EU funding, expected to raise additional €2.1bn of public and private investment
>
> Focus Area: Improve the development and uptake of AI.

The European Fund for Strategic Investment (EFSI)[95], on the other hand, addresses a more general lack of investment within the EU. It has a budget of €33.5bn (€7.5bn EIB capital) and expects to raise an additional €500bn by 2020. In 2017, the fund reached a total investment of €225.3bn.[96] However, these investments are not solely focused on the digital sector (which encompasses AI for the fund's purposes). The investments also support other strategic areas within the EU's economy, including transport, energy, education and research.

The EFSI was launched by the European Investment Bank (EIB) and the European Investment Fund (EIF)[97] alongside the European Commission and forms part of the Investment

Plan for Europe: the Juncker Plan.[98] It is built on the recognition that without investment in innovation, Europe will struggle to remain competitive. The Investment Plan for Europe has three core pillars. It identifies (i) a need to improve the regulatory environment, (ii) a need to support the European investment environment, and (iii) a need for a fund to financially support innovation. The European Investment Advisory Hub[99] and the European Union Investment Project Portal[100] support the Juncker Plan.

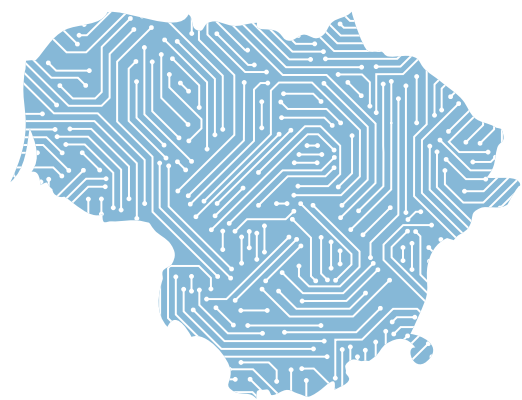

> **European Fund for Strategic Investment (EFSI)**
>
> Established: 2015
>
> Lead by: European Investment Bank, European Investment Fund, European Commission
>
> Budget: €26bn EU budget, €7.5bn from EIB; aims to unlock additional investment of €500bn until 2020
>
> Investment in 2017: €225.3bn
>
> Focus Area: Sectors of "key importance" to Europe e.g. energy, transport, digital, education, research, development and innovation.

The European Commission further commits itself to supporting 'breakthrough' innovation, as well as business ideas, through the European Innovation Council.[101] To that end, the European Innovation Council has a total of €2.7bn available (2018-2020) for startups and small companies to help them scale up. The European Innovation Council is not a fund exclusive for AI, but rather for any "highly-risky"[102] technology or business idea.

In terms of funding for research and innovation, the European Commission, through Horizon 2020,[103] will spend €1.5bn on AI between 2018-2020. This investment is expected to trigger an additional €2.5bn from existing Public-Private Partnerships, totalling €4bn.[104] This would constitute a 70% increase in funding compared to 2014-2017. The European Commission forecasts that if the private sector and Member States invested alongside this, then total investment into AI research, development and innovation in the EU could reach €20bn by the end of 2020, with a similar amount each subsequent year.[105]

> **Horizon 2020**
>
> Timeframe: 2014-2020
>
> Lead by: European Commission
>
> Objective: Horizon 2020 is the EU's biggest research and innovation programme with a total funding of around €80bn.
>
> AI specific funding (2018-2020): €2.5bn

Under the next Multiannual Financial Framework (MFF), funding for AI research will be made available by Horizon Europe, the successor to the Horizon 2020 programme, and the Digital Europe programme.[106] The Digital Europe programme will address areas such as AI, high-performance computing (HPC), digital skills and cybersecurity. The total amount currently proposed to be spent on AI, without matched funding from other sources, is €2.5bn. Other proposed funding includes €2bn for cybersecurity, €2.7bn for HPC and €0.7bn for advanced digital skills development.

The previous paragraphs outline several funding instruments as well as expected financial commitments. While these demonstrate acute awareness of the gaps and are designed to address them, it remains to be seen whether they will be sufficient, swift and targeted enough to make a meaningful impact.

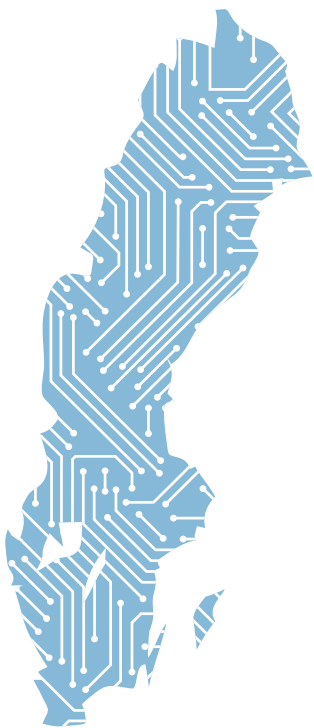

## Section (b): Talent creation

Section (b) focuses on the overlapping areas of brain drain and talent creation. After an introductory note on brain drain in the EU, it focuses on the EU's measures for talent creation.

Brain drain refers to a phenomenon where qualified workers leave a country or institution, often to receive a more competitive salary with better perks elsewhere. With an ongoing global competition for a limited supply of AI talent,[108] brain drain is not a phenomenon unique to the EU.[109] Irrespectively, the EU suffers from the migration of European academics to international companies and organisations within Europe as well as to other continents.[110,111] While the EU is host to many noteworthy academic institutions, salaries are often not competitive enough to keep researchers in academia and teach the next generation.[112,113] In addition to thereby minimising the potential future talent pool, many European researchers end up working for non-European owned AI companies.[114,115,116,117] So although they have not directly moved abroad they are now working for a non-European company as opposed to supporting European entreprises. This broadly poses three problems: how to keep talent in the EU (directly or indirectly), how to attract talent to the EU and how to ensure that enough talent is educated to satisfy continuous demand.

Several national strategy documents such as the French *Donner un sens à l'intelligence artificielle*,[118] the UK's *AI Sector Deal*,[119] or Finland's *Finland's Age of Artificial Intelligence*[120] recommend specific steps to tackle these issues. A recurrent suggestion is to increase the number and attraction of academic research careers, through increased funding as well as an increase in the number of PhD positions. For example, in order to increase the attraction of research careers, *Donner un sens à l'intelligence artificielle* suggests a doubling of early-career researchers' salaries, general salary top ups, a reduction in administrative formalities and a strong focus on interdisciplinarity research within and between institutes. Another suggestion is to increase the number of AI talent moving to the country in question, i.e. facilitating talent attraction.[121] The *AI Sector Deal*, for example, proposes a doubling of the number of Tier 1 visas (Exceptional Talent)[122,123] and a change to the immigration rules for settlement for Tier 1 visa holders. On the other hand, Finland's *Age of Artificial Intelligence* suggest the creation of an environment that appeals to the families of AI talent, for example, through access to international schools, employment opportunities for spouses, day care centres and an easy immigration process. Similarly, the European Commission notes the importance of "creating an attractive environment"[124] for talent attraction and retention.

Beyond Member State initiatives, the EU more broadly acknowledges that it may "risk losing out on the opportunities offered by AI, facing a brain drain and being a consumer of solutions developed elsewhere".[125] Yet, EU initiatives mainly focus on talent creation with the following paragraphs briefly mapping three areas: training and reskilling, traineeships and national strategies.

The EU introduced measures to address training and reskilling in the digital sector[126] with the Digital Skills and Jobs Coalition.[127,128] The intention of this Coalition is to endow citizens with the necessary digital skills to navigate society, their current as well as future jobs. It proposes suitable measures for four groups: broad society, labour force (upskilling, reskilling, jobseekers, career advice etc.), ICT professionals, and education (teaching, learning, lifelong learning, educating teachers). A specific milestone of the Digital Skills and Jobs Coalition is to train 1 million young unemployed people for vacancies in the digital sector by 2020.

Currently, there are ±350,000 vacancies for digital talent[129] in Europe with few people able to fill them. To further close this gap, the European Social Fund,[130] will support Member States with €2.3bn to use for digital skills development. This will be complemented by the recently launched

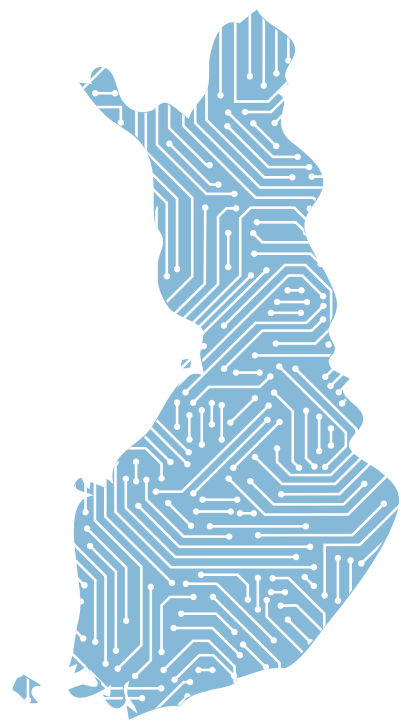

Digital Opportunity Traineeship.[131] The Digital Opportunity Traineeship is funded by Horizon 2020 and Erasmus+. It offers 6000 students and recent graduates funding for traineeships between 2018-2020.

Finally, as part of the New Skills Agenda for Europe,[132] the European Commission asked Member States to develop national digital skills strategies by mid-2017, alongside suitable implementation measures. It further issued recommendations to Member States on improving lifelong learning, digital literacy and numeracy. Consistent with the EU's vision of itself as a leader in 'ethical AI', the European Commission also clearly states that training courses and programmes on new technologies such as AI should incorporate ethical considerations.[133]

## Section (c): Putting the pieces back together

It is likely that collaboration between the Member States, beyond an alignment in AI strategies, would further strengthen the EU's position in the global AI ecosystem. The following paragraphs introduce suitable existing collaborations and discuss future opportunities in the space.

### (i) Large-scale collaboration

Subsections (i.a), (i.b) and (i.c) explore infrastructural collaborations, joint undertakings in hardware and proposals for pan-European AI laboratories.

### (i.a) Active collaboration: infrastructure

French multinational THALES leads the consortium that won the bid to develop a European AI-on-demand-platform.[134] The project, AI4EU,[135] is funded with €20m until 2021. Its aim is to create a collaborative AI platform that integrates the entire European AI ecosystem and 'democratises' access to AI.

According to the call for proposals, the final AI-on-demand platform is expected to compile and provide expertise, algorithms and other tools in an easily accessible format. It is presumed that such a repository will benefit businesses and sectors that do not yet have access to AI or do not have the relevant knowledge to implement AI.

Before that, however, AI4EU will develop eight industry driven AI pilots to explore the value of such a larger AI-on-demand platform as a "technological innovation tool". These pilots will focus on the use of AI for: citizens, robotics, industry applications, healthcare, media, agriculture, IoT and cybersecurity. Furthermore, the AI4EU project will establish an Ethics Observatory tasked with monitoring the adherence to human-centric values and create a Foundation managing concerns surrounding sustainability. Lessons learnt from these will feed into the Strategic Research and Innovation Agenda for Europe. The Agenda will further build on ongoing initiatives and strategies such as the robotics Public-Private Partnership,[136] and the cybersecurity Research Public-Private Partnership.[137]

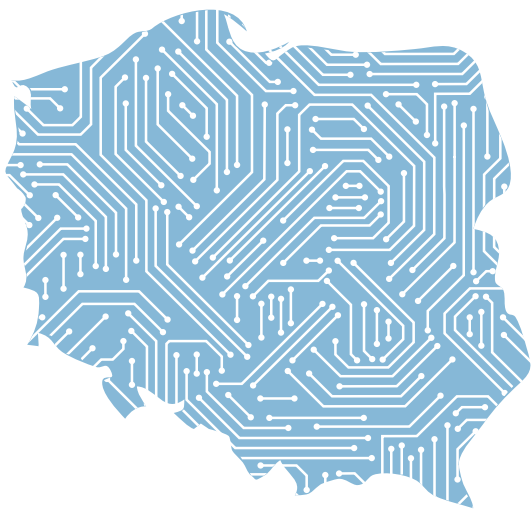

> **AI4EU (AI-on-demand platform)**[138]
>
> Budget: €20m
>
> Consortium: led by Thales, 79 partners in 21 countries are already participating in the project (full list available on the project's website)[139]
>
> Objective: "1. serve as a central point to gather and provide access to AI-related knowledge, algorithms and tools; 2. support potential users of AI in order to facilitate the integration of AI into applications; 3. facilitate the interaction with existing data portals needed for AI algorithms, and resources, such as HPC or cloud computing, and support interoperability."[140]

Another infrastructural collaboration is found in the Digital Innovation Hubs (DIH)[141] network which is part of the Digital Single Market package[142] and the Digitise European Industry[143] effort.

There are currently ~450 DIH, each acting as a regional first point-of-contact and provider of a full range of services for SMEs and other businesses. All Hubs are interconnected, together creating a distributed pan-European network of resources. Individual Hubs are made up of organisations such as local universities, incubators/accelerators and industry associations. As a provider, a Hub grants access to sector specific expertise and technology, supporting for example, companies in their testing and experimentation phases. Users of Hubs can also gain access to competence centres (e.g. providing further technological infrastructure), financial and business advice.[144] While the current network predominantly focuses on robotics, the European Commission's Communication on AI suggested that it could be used to support a "first series of testing and experimentation infrastructures for AI products and services".[145]

Bearing this in mind, the European Commission, alongside PwC, Carsa and Innovalia, is currently calling for 30 Digital Innovation Hubs in the field of AI.[146] These AI-specific Hubs are expected to kickstart a network for collaboration across the EU, as called upon by open letters from the research community such as CLAIRE. In addition, these new AI-specific Hubs are expected to be involved in training programmes, and to "provide evidence for the preparation of relevant European policies".[147]

**Digital Innovation Hubs**

Location: ~450 DIHs across Europe

Budget: €500m over 5 years from the Horizon 2020 framework programme

Objective: "[...] help European Industry, small or large, high-tech or not, to grasp the digital opportunities"[148]

The following paragraphs introduce different areas of collaboration relevant to AI. In particular, focus is given to SPARC, the Electronic Components and Systems Joint Undertaking and the European High-Performance Computing Joint Undertaking.

**(i.b) Active collaboration: Public-Private Partnerships and hardware**

SPARC[149] is a robotics Public-Private Partnership (PPP) between the European Commission, industry and academia. It is among the largest civilian funded robotics innovation programmes in the world, with €700m from the European Commission and €2.1bn from industry. SPARC primarily aims to strengthen Europe's competitiveness in the field of industrial robotics, where Europe holds a ~32% world market share.[150] Its broader mission is to redistribute the benefits derived from robotics towards the wider society and reinvest into Europe's economy. To that end, it also promotes the growth of the robotics value chain through public engagement, e.g. the European Robotics week and competitions, e.g. the European Robotics League.

SPARC also consults the European Commission on shifting developments in research, development and innovation through the Strategic Research Agenda (SRA)[151] and the Multi-Annual Roadmap (MAR).[152] Both documents provide input into the strategic next steps for the European Commission for the twin focus areas of funding and emerging societal challenges associated with robotics.

Another, more directly relevant Public-Private Partnership is the Electronic Components and Systems Joint Undertaking (ECSEL).[154] Electronic components and systems are used in most smart devices, e.g. smartphones, smart energy grid or smart cards and can contribute to the development of a variety of newer technologies, such as neuromorphic chips. The ECSEL PPP, similar to the SPARC PPP within the field of robotics, aims to establish Europe as a global leader in the electronic components and systems industry while supporting the existing ecosystem. ECSEL funds research, development and innovation projects through open calls. In addition, it is in the process of strengthening existing clusters and creating new ones across Europe, ensuring that actors can access the relevant technical infrastructure for the manufacturing and design of electronic components and systems.

The third relevant initiative is the European High-Performance Computing Joint Undertaking (EuroHPC JU),[159] established in November 2018. It is a timely move for the EU as high-performance computing has the potential to significantly improve EU competitiveness. One facet of the current lack of supercomputers in the EU, is that researcher often face the need to process their data outside of Europe, possibly under disadvantageous conditions. As a first step towards remedying this situation, the EuroHPC JU will develop a "a clear strategy for innovation procurement of exascale machines based on competitive European technologies".[160] In addition, it will aid Member States and other European countries in the coordination of their strategies and investments, with the expectation that this will lead to an increase in the development of supercomputers within the European market.

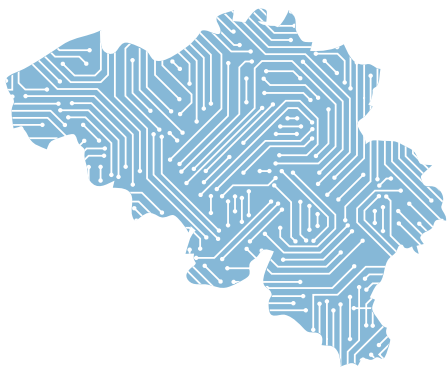

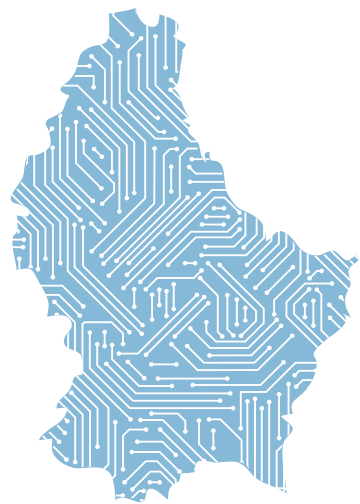

> **European High-Performance Computing Joint Undertaking**
>
> Established: November 2018
>
> Budget: €486m EU funding (Horizon 2020 and Connecting Europe Facility), total funding of around €1bn, including additional private funding of > €400m[161]
>
> Location: Luxembourg, Luxembourg
>
> Members: EU via the European Commission; Austria, Belgium, Bulgaria, Croatia, Czech Republic, Denmark, Estonia, Finland, France, Germany, Greece, Hungary, Ireland, Italy, Latvia, Lithuania, Luxembourg, Netherlands, Norway, Poland, Portugal, Romania, Slovakia, Slovenia and Spain. (Member States and associated countries that signed the declaration and joined since);[162] European Technology Platform for High Performance Computing (ETP4HPC), Big Data Value (BDVA) and other relevant industry and stakeholder representatives
>
> Objective: "[..] acquiring and providing a world-class pre-exascale[163] supercomputing infrastructure to Europe's scientific and industrial users, matching their demanding application requirements by 2020; developing exascale supercomputers based on competitive EU technology that the Joint Undertaking could acquire around 2022/2023, and that would be ranking among the top three places in the world."[164]

**(i.c) Proposals for collaboration: pan-European AI laboratories**

Many AI researchers and experts are taking note of the EU's demonstrated strength in large scientific research collaborations,[165] but also of its limited success in transferring this expertise to the field of AI research and development. The latter has been a growing topic of concern in the community, notably expressed through a call to establish a CERN for AI, and two open letters: the European Lab for Learning and Intelligent Systems and the Confederation of Laboratories for Artificial Intelligence Research in Europe. What follows is a brief overview of these three proposals.

As opposed to the latter two, the call for a 'CERN for AI', modelled off the European Organisation for Nuclear Research (CERN), originates from a single advocate, Prof. Gary Marcus. An international collaboration on AI of a scale equal to CERN would ensure that research results are shared globally,[166] accommodate a high number of interdisciplinary scientists[167] and benefit from significant funding for ambitious foundational research, he argues.

> **CERN for AI**
>
> Announcement date: 2017
>
> Proponent: Prof. Gary Marcus
>
> Idea: build a pan-European organisation similar to the European Organization for Nuclear Research (CERN), with equal financial backing and buy-in from a majority of EU Member States
>
> Status: unclear

In a larger effort, a group of prominent researchers called for the establishment of a European Lab for Learning and Intelligent Systems (ELLIS).[168] Their open letter argues that while the EU benefits from strong academic research centers, their current distribution across Member States may cause a problematic fragmentation for the EU in the long run. It further notes that a suitable course of action would be to connect capabilities into a cross-national European AI Lab, supported by sufficient funding from participating countries.[169]

In pursuit of this plan, ELLIS recently announced the "formation of its professional association"[170] tasked with the management and the creation of the structures needed for the envisaged goal. In addition, ELLIS will push for a pan-European PhD programme and focus on industrial engagement. In terms of research, it aims to hone in on machine learning methods.

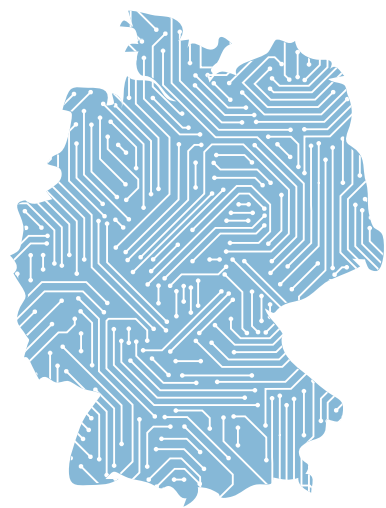

Similar to the rest, it argues that collaboration and coordination within academia and research could contribute to a strengthening of the EU AI ecosystem. To that end, it suggests the establishment of a distributed confederation of AI laboratories with a wide-range of application areas as well as a considerable focus on AI research for social impact. Unlike others, however, CLAIRE emphasises the importance of the development of trustworthy AI. Since the open letter got published, supporters of ELLIS and CLAIRE agreed to "jointly push for common infrastructure, including compute resources "at Google scale"".[173]

The intent to connect AI labs[174] also features in recent policy publications such as Germany's Eckpunkte der Bundesregierung für eine Strategie Künstliche Intelligenz.[175]

> **European Lab for Learning and Intelligent Systems**
>
> Announcement date: 2018
>
> Proponents: Prof. Zoubin Ghahramani, Chief Scientist at Uber, and Bernard Schölkopf, director at the Max Planck Institute for Intelligent Systems in Tübingen, Germany
>
> Idea: "1. we want the best basic research to be performed in Europe, to enable Europe to shape how machine learning and modern AI change the world, and 2. we want to have economic impact and create jobs in Europe, and believe this is achieved by outstanding and free basic research, independent of industry interests."[171]
>
> Status: formally announced "the formation of its professional association that will undertake the organization and building of the intellectual and physical structures of ELLIS"[172] in December 2018.

The third and most recent proposal in this same direction is calling for a Confederation of Laboratories for Artificial Intelligence Research in Europe (CLAIRE).

> **Confederation of Laboratories for Artificial Intelligence Research in Europe**
>
> Announcement date: June 2018
>
> Signatories: over 1900 supporters[176]
>
> Objective: "We call for a vision that aims to (1) have European research and innovation in artificial intelligence be amongst the best in the world, that (2) encompasses all of AI and all of Europe, and that (3) has a strong focus on human-centred AI"[177]
>
> Status: first symposium in September 2018[178]

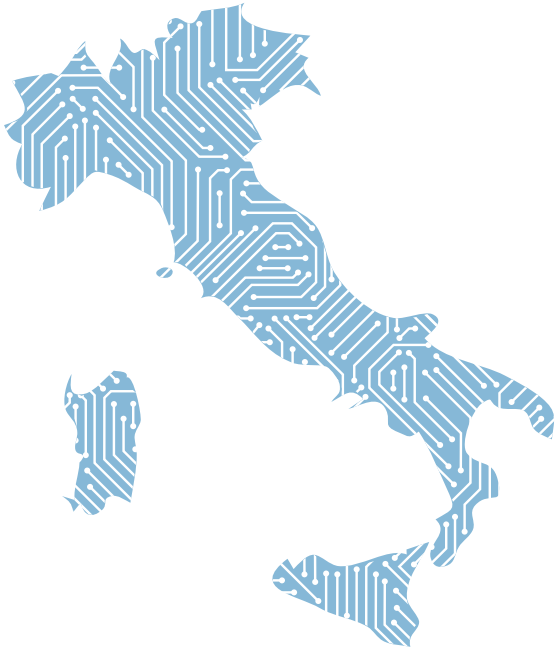

All three calls to action point towards a significant deficit within the EU, as identified by the academic community. This deficit will likely need to be addressed in a timely manner if Europe wishes to remain competitive and strengthen its leadership in AI.

**(ii) Granular collaboration**

Whereas the previous section introduced large-scale collaborations such as existing frameworks, resources and ongoing proposals for an EU AI laboratory, the following paragraphs focus on smaller scale initiatives between individual groups and actors within Member States.

The Joint European Disruptive Initiative (JEDI)[179,180] is an upcoming initiative by actors within France and Germany. JEDI hopes to become a 'European idea-factory' that thinks about and funds long-term and high-risk research, an area in which it sees the EU lagging behind.[181] JEDI is currently assembling an expert committee which it expects to identify technologies that would enable the European 'deep tech ecosystem' to thrive, if funded by JEDI. As a first step, JEDI is expected to launch 'TechChallenges' under the guidance of business executives, academic researchers and political advisors,[182] with a focus on AI, cybersecurity, biotechnologies and nanotechnologies.[183]

A conceptually similar idea, that of an EU Agency on Disruptive Innovation, was proposed by President Macron in 2017.[184,185] The importance of such an Agency for Disruptive Innovation is also emphasized in the German Eckpunkte der Bundesregierung für eine Strategie Künstliche Intelligenz[186] published by the Chancellor's Office in July 2018. In fact, the German 9-Punkte-Plan[187] issued by the Bundesverband für Künstliche Intelligenz (Federal Association for Artificial Intelligence) directly references JEDI in this regard.

> **Joint European Disruptive Initiative**
>
> Budget: provisional budget expected ~ €235m; to expand to €1bn per year
>
> Established: 2018
>
> Members: lead by André Loesekrug-Pietri, Jean-Paul Palomeros and Didier Schmitt;[188] conceptually similar ideas have been endorsed by President Macron and Chancellor Merkel as well as in previous meetings hosted by the European Commission[189]
>
> Objective: It hopes to build an agency that identifies and finances moonshot-type projects in Europe.

During the 2018 AI for Humanity Summit[190] in France, president Macron presented the Paris Artificial Intelligence Research Institute (PRAIRIE) initiative.[191] With a more national regard than ELLIS and CLAIRE, this initiative looks to strengthen the connection of French AI institutes with other global leaders. The PRAIRIE initiative is currently in its inaugural phase and will be an expansion of a collaborative agreement between several relevant French institutes with leading global AI research centres (e.g. MILA in Montreal, the Max Planck Institute in Tübingen, Germany, and CIIRC in the Czechia). It expects to provide training for researchers at all career stages, to support novel research and knowledge transfer as well as to establish close-knit channels between academia and industry. The latter will be supported by a number of industry collaborators forming part of the PRAIRIE's network. Although this initiative outlines both a more nationalistic and global focus than those proposed by ELLIS and CLAIRE, it could still act as a suitable blueprint for a large scale European distributed AI lab.[192]

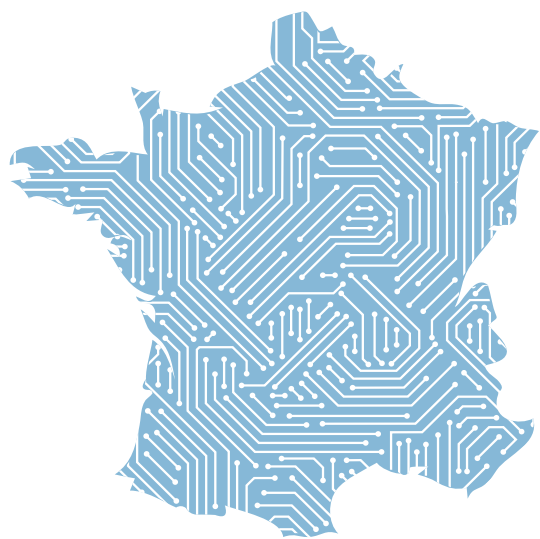

**Paris Artificial Intelligence Research Institute**

Budget: unclear

Established: 2018

Location: Paris, France

Objective: "The partners in PRAIRIE Institute (Paris Artificial Intelligence Research Institute) are pursuing three goals:

1. To make a significant contribution to driving progress in fundamental knowledge in AI freely distributed among the international scientific community;

2. To take part in solving concrete problems with major application-related impact;

3. To contribute to the training in the field of AI.

The five-year objective is to bring together AI scientific and industrial leaders and make the PRAIRIE Institute a world leader in AI."[193]

Members: French collaborators (academia and industry): CNRS, Inria and PSL University, together with Amazon, Criteo, Facebook, Faurecia, Google, Microsoft, NAVER LABS, Nokia Bell Labs, PSA Group, SUEZ and Valeo; EU collaborators: the Max Planck institute[194] in Tübingen, the CIIRC[195] (Czech Institute of Informatics, Robotics and Cybernetics) and the Alan Turing Institute[196] in London; International collaborators: Center for Data Science at NYU[197], the artificial intelligence laboratory of UC Berkeley (BAIR),[198] the Robotics institute[199] at Carnegie-Mellon University in Pittsburgh, MILA[200] in Montreal.

As mentioned previously, the EU benefits from a reputable and thriving academic AI community[201] and many community members constitute the driving force behind CLAIRE and ELLIS. The European Association for Artificial Intelligence (EurAI),[202] acts as the representative body of this community. Its mission is to encourage AI research and applications. To that end, it provides awards, grants and organises summer schools for researchers, in addition to sponsoring relevant Masters degrees as well as conferences such as IJCAI-ECAI. It is a valuable player within the EU's AI ecosystem by virtue of connecting and strengthening the existing community and nurturing upcoming talent.[203] Given EurAI's reach, engagement and activities, it could feasibly play a crucial role in a pan-European AI lab network and support the strengthening of a cooperative research environment within the EU.

## Conclusion

The above sections focus on some of the building blocks needed to support an EU AI strategy: (a) funding, (b) talent creation and (c) collaboration.

Section (a) outlines funding mechanisms such as VentureEU and the European Fund for Strategic Investment as well as other sources of funding for research, development and innovation (e.g. Horizon 2020, the European Innovation Council pilot and the Digital Europe Programme). It concludes that the EU is addressing past shortcomings in a variety of areas such as VC funding, however, it is unclear whether new initiatives will be successful or swift enough to achieve a change in direction. Section (b) addresses brain drain and talent creation. While the EU supports talent creation through education, upskilling and reskilling (e.g. the Digital Opportunity Traineeship and the Digital Skills and Jobs Coalition), it appears to pay less attention to talent retainment and attraction.

Section (c) focuses on collaboration between a variety of large (e.g. EU institutions, Member States) and small actors (e.g. groups within Member States). Subsection (i) outlines existing pan-European frameworks and resources (e.g. DIH, SPARC, ECSEL, EuroHPC JU). It then proceeds to examine three ongoing proposals for a pan-European AI lab which could feasibly build upon and use these existing resources: CERN for AI, ELLIS and CLAIRE. Subsection (ii) introduces more granular initiatives such as JEDI and PRAIRIE. Although there are several promising collaborations within the EU, only further reviews will clarify whether these are necessary and sufficient for the EU to strengthen its leadership in AI.

# Summary

This report explores some of the necessary and existing elements for AI leadership in the EU. In doing so, it provides an introductory overview of the EU's AI ecosystem by surveying relevant policy and strategy documents, regulations, projects and initiatives.

The elements explored are strategy and vision, funding, talent creation and collaboration.

Overall, the EU may have an opportunity to distinguish itself from other nations with its strategy and vision, if the current focus on 'ethical AI' is expanded, clarified and implemented. Depending on how this effort turns out, it could contribute to the EU's longer-term competitive advantage. New initiatives and pilot programmes address historic gaps in VC funding, alongside billions dedicated towards AI R&D during the remaining Horizon 2020 framework programme and the upcoming Digital Europe Programme at EU-level alone. This tackles some prior blind spots and creates a suitable stepping stone for the EU to go forwards. Unfortunately, actual impact of many funding initiatives will only be clear several years into their term. While brain drain is of concern to the EU, there are few successful counteractions momentarily identified and implemented. Instead, the EU is undertaking an array of initiatives to support talent creation at all levels: education, reskilling and upskilling. Finally, the EU benefits from various large scale and smaller projects of collaborative nature in areas such as hardware, research and infrastructure that could strengthen the EU as a singular actor.

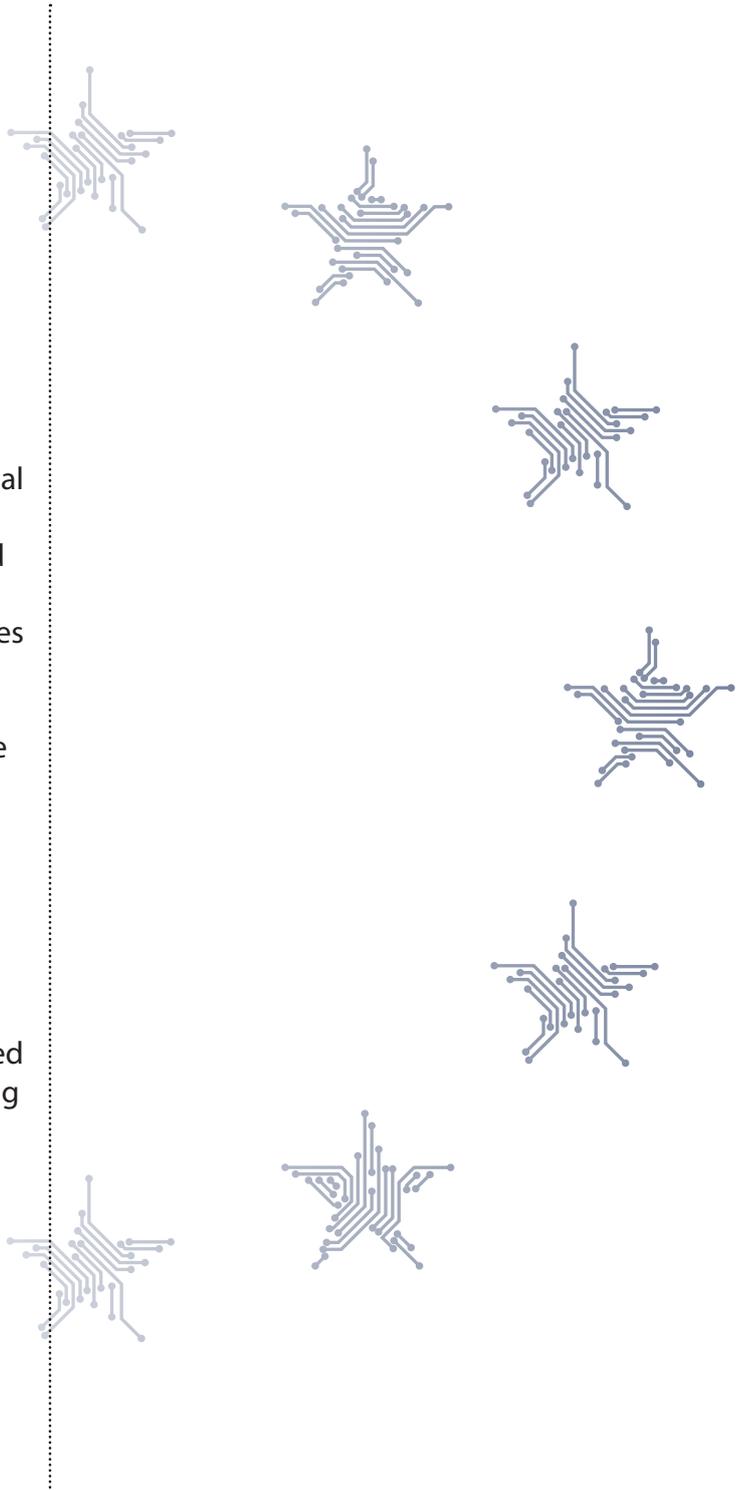

# Annex

Ongoing models for large-scale collaboration

In the broader field of science and technology, two prominent contenders for established pan-European collaborations are the European Organization for Nuclear Research (CERN) and the Human-Brain Project (HBP). CERN is "Europe's first joint venture"[234] and the Human-Brain Project is a European flagship project, which means that it is one of the EU's "visionary, science-driven, large-scale research initiatives addressing grand Scientific and Technological (S&T) challenges".[235] CERN and the Human-Brain Project receive significant funding from the EU and additional financial backing is provided by each participating Member State. CERN and the HBP are significant institutions for Europe, acting as signposts for the EU's capability to undertake large-scale collaborative research and could act as inspiration for what European AI projects or labs of similar scope may look like.

CERN's main research aim is to study particle physics. It is most commonly associated with the discovery of the Higgs boson particle. Based on a single site, CERN currently has 22 involved countries, alongside several Associated Countries and organisations listed as observers.

**European Organization for Nuclear Research**

Established: 1954

Location: Geneva, Switzerland

Members (countries): Belgium, Denmark, France, Germany, Greece, Italy, Netherlands, Norway, Sweden, Switzerland, United Kingdom, Austria, Spain, Portugal, Finland, Poland, Czechia, Hungary, Bulgaria, Israel, Romania, Slovak Republic

Associate Members States in pre-stage to membership: Cyprus, Serbia, Slovenia; Associated Member States: India, Lithuania, Pakistan, Turkey, Ukraine[236]

Observer Status: UNESCO, European Commission, the Joint Institute for Nuclear Research (JINR), Japan, Russia, United States

Staff: 2,633; 12,236 users (as of 31/12/2017)

Budget: CHF 1,148,2m (as of 13/03/18)

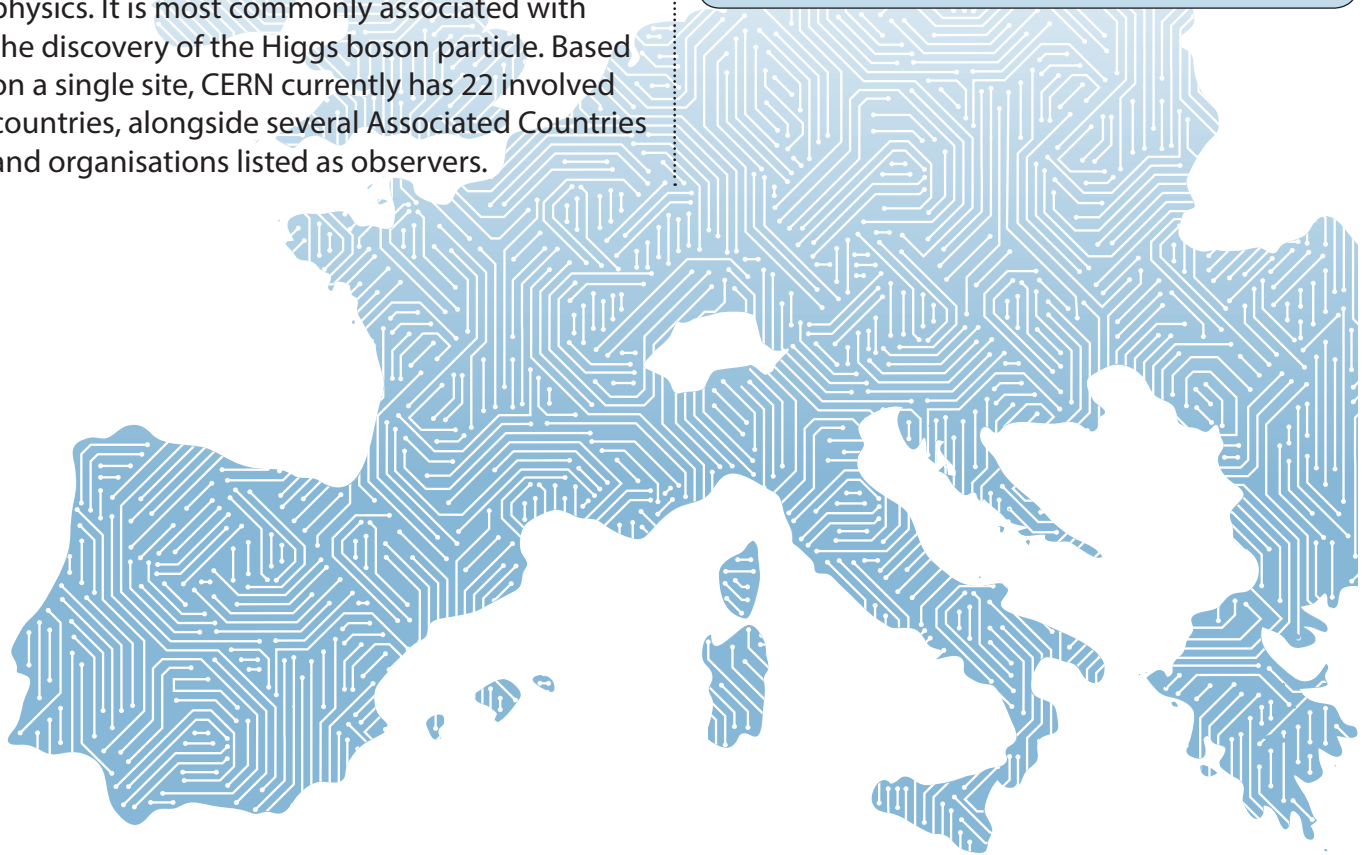

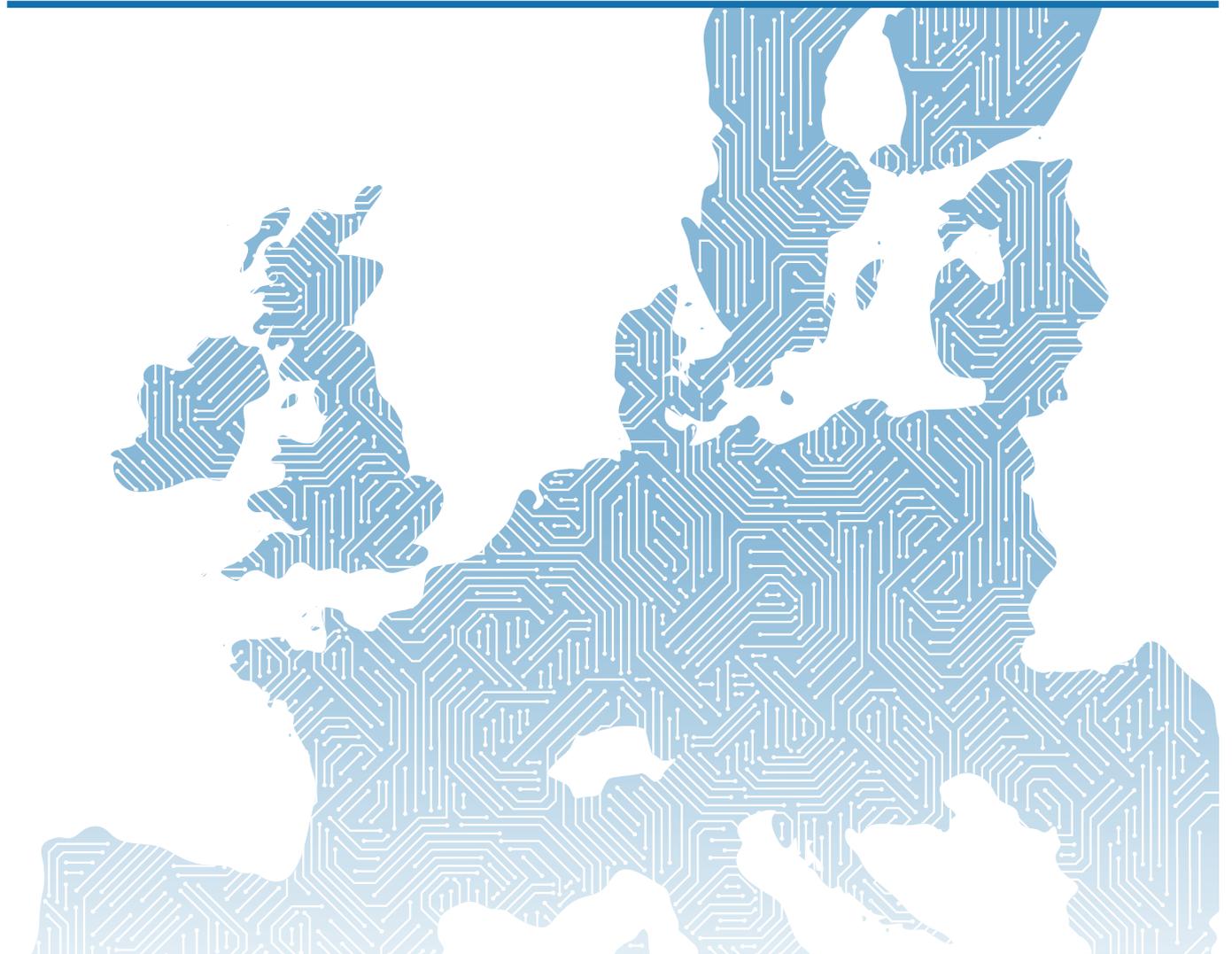

As of 2018, the Human-Brain Project is in its 5th year of operation, and is one of the largest scientific projects ever funded by the EU.[237] The project looks to advance the understanding of the human brain through research strands covering, for example, neuroscience, medicine and computing, with dedicated ICT platforms for topics such as neuroinformatics, medical informatics and neuromorphic computing. Relevant to discussions in this report, the HBP recently launched a project entitled SHERPA.[238] SHERPA will explore the ethical and societal impacts of the work[239] undertaken at the HBP and is funded with €2.8m. Specifically, SHERPA will use case studies and future scenarios to clarify how human rights and ethics could be impacted by AI and big data. This work appears aligned with The Toronto Declaration,[240] which seeks to work from the existing framework of international human rights standards, and to apply these to the development and use of AI systems.

**Human-Brain Project**[241]

Established: 2013 (until 2023)

Location: Geneva, Switzerland

Members (countries): Austria, Belgium, Denmark, Finland, France, Germany, Greece, Hungary, Italy, Netherlands, Norway, Portugal, Slovenia, Spain, Sweden, Switzerland, Turkey, United Kingdom, Israel

Staff: ~444 full-time employees across Europe at 116 Universities and research centres

ICT Research Platforms: Neuroinformatics, Brain Simulation, High Performance Analytics and Computing, Medical Informatics, Neuromorphic Computing, Neurorobotics

Budget: €89m (04/2016-03/2018)

E60605D068AC28C22D&gwt=pay&asset-Type=opinion

# Footnotes

1. This report focuses on the EU rather than Europe, the continent. Respectively, the adjective 'European' refers to the political and economic union.

2. This report does not consider the implications of Brexit.

3. e.g. https://www.technologyreview.com/s/608112/who-is-winning-the-ai-race/; https://www.wired.co.uk/article/why-china-will-win-the-global-battle-for-ai-dominance, https://www.forbes.com/sites/forbestechcouncil/2017/12/05/these-seven-countries-are-in-a-race-to-rule-the-world-with-ai/#cc695f4c245f;

4. https://www.politico.eu/article/opinion-europes-ai-delusion/

5. https://www.merics.org/en/blog/europes-ai-strategy-no-match-chinas-drive-global-dominance

6. https://www.politico.eu/article/merkel-artificial-intelligence-warns-brain-drain-to-foreign-tech-companies/

7. https://www.nrc.nl/nieuws/2018/08/27/nederland-kampt-met-ai-braindrain-a1614393

8. See: https://ec.europa.eu/digital-single-market/en/news/european-artificial-intelligence-landscape

9. http://data.consilium.europa.eu/doc/document/ST-8507-2018-INIT/en/pdf

10. https://www.entrepreneur.com/article/313692

11. The total number expected, taking into account existing Public-Private Partnerships, is €4bn.

12. This number excludes any matched or industry funding as well as funding on Member State level.

13. http://europa.eu/rapid/press-release_IP-18-2763_en.htm

14. https://www.recode.net/2016/7/18/12213472/softbank-buying-arm-chip-design

15. https://www.bloomberg.com/news/articles/2017-03-08/midea-eyes-top-spot-for-kuka-in-china-s-booming-robot-market

16. https://techcrunch.com/2016/06/20/twitter-is-buying-magic-pony-technology-which-uses-neural-networks-to-improve-images/

17. https://www.politico.eu/article/opinion-europes-ai-delusion/

18. https://www.merics.org/en/blog/europes-ai-strategy-no-match-chinas-drive-global-dominance

19. https://www.politico.eu/article/merkel-artificial-intelligence-warns-brain-drain-to-foreign-tech-companies/

20. https://www.nrc.nl/nieuws/2018/08/27/nederland-kampt-met-ai-braindrain-a1614393

21. See: https://ec.europa.eu/digital-single-market/en/news/european-artificial-intelligence-landscape.

22. For the EU's purposes this includes, but is not limited to, the development, application and deployment of AI.

23. Full title: "Artificial intelligence – The consequences of artificial intelligence on the (digital) single market, production, consumption, employment and society"

24. https://www.bmbf.de/files/180718%20Eckpunkte_KI-Strategie%20final%20Layout.pdf; no official English translation. The title translates to Cornerstones of the Federal Government for an Artificial Intelligence Strategy.

25. http://julkaisut.valtioneuvosto.fi/bitstream/handle/10024/160391/TEMrap_47_2017_verkkojulkaisu.pdf?sequence=1&isAllowed=y

26. https://www.aiforhumanity.fr/pdfs/9782111457089_Rapport_Villani_accessible.pdf

27. It is not within the scope of this report to examine the whole landscape relevant to any form of AI leadership.

28. http://europa.eu/rapid/press-release_IP-18-6689_en.htm

29. https://ec.europa.eu/digital-single-market/en/news/eu-member-states-sign-cooperate-artificial-intelligence

30. A declaration is a legally non-binding document. It is an official statement outlining the intent of its signatories.

31. See: https://ec.europa.eu/digital-single-market/en/news/eu-member-states-sign-cooperate-artificial-intelligence.

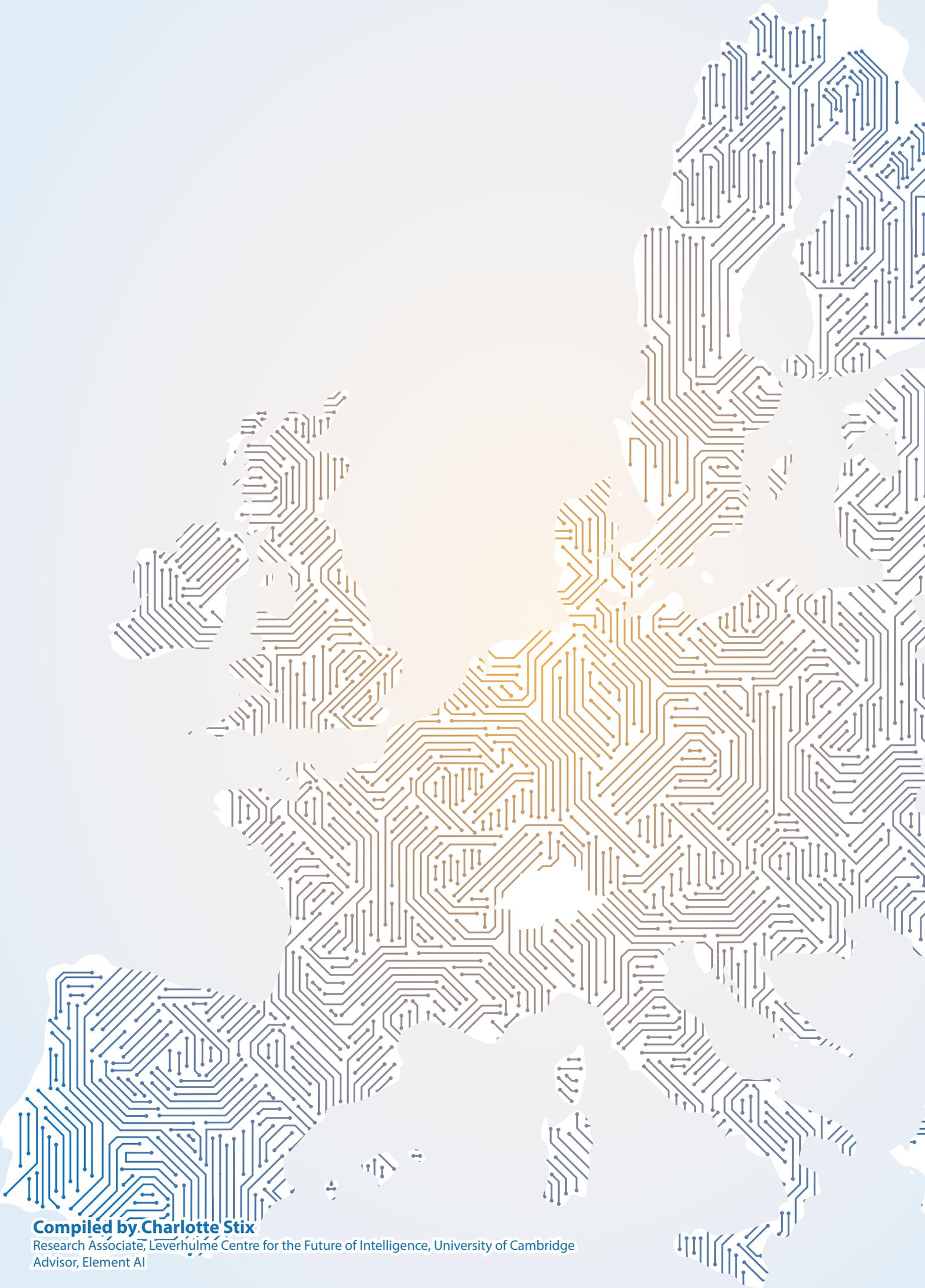

**Compiled by Charlotte Stix**
Research Associate, Leverhulme Centre for the Future of Intelligence, University of Cambridge
Advisor, Element AI